\documentclass[longauth]{aa} 
\usepackage{graphicx}
\usepackage{tabularx}
\usepackage{multirow}
\usepackage[colorlinks=true, linkcolor=blue, citecolor=blue]{hyperref}
\usepackage{gensymb}
\usepackage{color}
\usepackage{txfonts}
\usepackage{natbib}
\usepackage{amsmath}
\usepackage{ulem}
\usepackage[flushleft]{threeparttable}
\usepackage[dvipsnames]{xcolor}
\usepackage{rotating}
\usepackage{CJKutf8}

\newcommand{\nii}{[N\,\textsc{ii}]}

\newcommand{\sii}{[S\,\textsc{ii}]}

\begin{document} 

\titlerunning{Revised Redshift of S1}
\authorrunning{Xiao et al.}
   \title{
 From Grism to IFU: Revising the Redshift and Nature of the Massive Dusty Galaxy S1 with JWST and ALMA}

\author{Mengyuan Xiao\inst{\ref{unige}}\thanks{E-mail: mengyuan.xiao@unige.ch}
\and Longji Bing\inst{\ref{Sussex}}
\and Gabriel Brammer\inst{\ref{dawn},\ref{NBI}} 
\and Pascal A. Oesch\inst{\ref{unige},\ref{dawn},\ref{NBI}}
\and David Elbaz\inst{\ref{cea}}
\and Rui Marques-Chaves\inst{\ref{unige}} 
\and Miroslava Dessauges-Zavadsky\inst{\ref{unige}} 
\and Benjamin Magnelli\inst{\ref{cea}} 
\and Rychard Bouwens\inst{\ref{leiden}} 
\and Emanuele Daddi\inst{\ref{cea}} 
\and Maximilien Franco\inst{\ref{cea}} 
\and Qiusheng Gu\inst{\ref{nju1},\ref{nju2}} 
\and Thomas Herard-Demanche\inst{\ref{leiden}} 
\and Garth Illingworth\inst{\ref{ucsc}} 
\and Ivo Labbe\inst{\ref{Swinburne}}
\and Danilo Marchesini\inst{\ref{Tufts}}
\and Jorryt Matthee\inst{\ref{Austria}} 
\and Romain A. Meyer\inst{\ref{unige}} 
\and Rohan P. Naidu\inst{\ref{mit}} 
\and Irene Shivaei \inst{\ref{cab}} 
\and Pieter van Dokkum\inst{\ref{yale}} 
\and Andrea Weibel\inst{\ref{unige}} 
\and Christina C. Williams\inst{\ref{Arizona},\ref{NSF}} 
\and Stijn Wuyts\inst{\ref{bath}} 
}

\institute{Department of Astronomy, University of Geneva, Chemin Pegasi 51, 1290 Versoix, Switzerland\label{unige}
\and Astronomy Centre, University of Sussex, Falmer, Brighton BN1 9QH, UK  \label{Sussex}
\and Cosmic Dawn Center (DAWN), Denmark \label{dawn}
\and  Niels Bohr Institute, University of Copenhagen, Jagtvej 128, K\o benhavn N, DK-2200, Denmark \label{NBI}
\and Universit{\'e} Paris-Saclay, Universit{\'e} Paris Cit{\'e}, CEA, CNRS, AIM, 91191, Gif-sur-Yvette, France \label{cea}
\and Leiden Observatory, Leiden University, PO Box 9500, 2300 RA Leiden, The Netherlands\label{leiden}
\and School of Astronomy and Space Science, Nanjing University, Nanjing, Jiangsu 210093, China\label{nju1}
\and Key Laboratory of Modern Astronomy and Astrophysics, Nanjing University, Ministry of Education, Nanjing 210093, China\label{nju2}
\and Department of Astronomy and Astrophysics, University of California, Santa Cruz, CA 95064, USA\label{ucsc}
\and Centre for Astrophysics and Supercomputing, Swinburne University of Technology, Melbourne, VIC 3122, Australia\label{Swinburne}
\and Department of Physics and Astronomy, Tufts University, 574 Boston Avenue, Medford, MA 02155, USA\label{Tufts}
\and Institute of Science and Technology Austria (ISTA), Am Campus 1, 3400 Klosterneuburg, Austria \label{Austria}
\and MIT Kavli Institute for Astrophysics and Space Research, 70 Vassar Street, Cambridge, MA 02139, USA \label{mit}
\and Centro de Astrobiología (CAB), CSIC-INTA, Carretera de Ajalvir km 4, Torrejón de Ardoz, E-28850, Madrid, Spain\label{cab}
\and Astronomy Department, Yale University, 52 Hillhouse Ave, New Haven, CT 06511, USA\label{yale}
\and Steward Observatory, University of Arizona, 933 N Cherry Avenue, Tucson, AZ 85721, USA \label{Arizona}
\and NSF National Optical-Infrared Astronomy Research Laboratory, 950 North Cherry Avenue, Tucson, AZ 85719, USA \label{NSF}
\and Department of Physics, University of Bath, Claverton Down, Bath, BA2 7AY, UK\label{bath}
}
\date{Received xxx; accepted xxx}

  \abstract
 {
We report a revised spectroscopic redshift for the dusty massive galaxy S1, which was previously inferred to have an exceptionally high baryon-to-star conversion efficiency from NIRCam slitless grism data at $z_{\rm grism}=5.58$.  
Our new JWST/NIRSpec IFU observations reveal multiple rest-frame optical and and near-infrared emission lines ([S\,\textsc{iii}]$\lambda9069$, [S\,\textsc{iii}]$\lambda9531$, He\,\textsc{i}~$\lambda1.083\,\mu$m, and Pa$\gamma$), yielding a secure spectroscopic redshift of $z_{\rm spec}=3.2439\pm0.0002$.  
We show that the earlier grism-based redshift was caused by contamination from a nearby galaxy whose dispersed spectral trace overlaps with that of S1. This highlights the known challenge of slitless spectroscopy, 
when only a single dispersion angle is observed and only a single emission feature is detected.  
 In addition, we present new ALMA 1\,mm continuum observations of S1, which robustly detect dust emission ($S_{\rm 1mm}=0.99\pm0.03$\,mJy) and show that the dust half-light radius ($R_{\rm e,1mm}=0.73\pm0.10$~kpc) is slightly smaller than the stellar size ($R_{\rm e, F444W} = 0.97\pm0.01$\,kpc) measured in JWST/F444W.  
Using the revised redshift and compiled multi-wavelength photometry, we update the UV-to-FIR spectral energy distribution (SED) of S1. We find that S1 is less extreme than previously inferred when assumed to be at $z_{\rm grism}=5.58$, but remains a very massive (log$M_{\star}/M_{\odot}$ $\sim10.6$), heavily obscured star-forming galaxy at $z_{\rm spec}=3.2439$. The updated SED modeling reveals S1 to be a very dust- and gas-rich system with a moderate star formation rate ($\rm SFR\sim70$ $M_\odot$ yr$^{-1}$) and a long gas depletion time ($\tau_{\rm dep} \sim 1.4$ Gyr), deviating from submillimeter galaxies (SMGs) and optically dark or faint galaxies (OFGs), but more closely resembling typical massive main-sequence galaxies at a similar redshift. We note that, although this revision reduces the number of ultra-massive galaxies reported in \citet{Xiao2024}, it does not alter the main conclusions of that work. 
Overall, our study clarifies the nature of S1 and underscores the importance of multi-line spectroscopic confirmation, slitless observations obtained at multiple position angles, and IFU data for robust redshift determination and physical interpretation of rare, massive galaxies in the early Universe.
\\
 }

   \maketitle
%

\section{Introduction}
The advent of JWST has transformed our understanding of massive galaxy formation in the early Universe, revealing a substantial population of massive/ultra-massive candidates at $z\gtrsim5$ \citep[e.g.,][]{Xiao2024, Carnall2024, Gentile2024, Weibel2024, Chworowsky2024,Xiao2025_spiral, Shuntov2025,deGraaff2025}.  
These galaxies often exhibit large stellar masses and, in many cases,  high infrared luminosities and significant dust obscuration \citep[e.g.,][]{Lagache2025,Bing2025}, raising important questions about the efficiency of baryon conversion into stars during the first billion years after the Big Bang \citep[e.g.,][]{Boylan-Kolchin2023}.  

Slitless NIRCam grism spectroscopy has played a central role in these discoveries by enabling wide-field, blind searches for emission-line galaxies \citep[e.g.,][]{Oesch2023, Meyer2025_cosmos3d}.  
In particular, the FRESCO and CONGRESS surveys have provided spectroscopic redshifts for thousands of sources in the GOODS fields \citep[e.g.,][]{Covelo-Paz2025_Ha, Meyer2024, Neufeld24, Zhang25_CONGRESS}, offering a powerful statistical view of galaxy evolution at high redshift (see also EIGER \citealt{Kashino23_EIGER}, ALT \citealt{Naidu24_ALT}, COSMOS-3D \citealt{Meyer2025_cosmos3d}, or the pure parallel surveys SAPPHIRES and POPPIES \citealt{Sun25_SAPPHIRES}).  
At the same time, a well-known challenge of slitless spectroscopy is contamination from overlapping spectral traces of neighbouring sources, which can complicate source attribution and line identification, especially in crowded fields and for single-line detections \citep[e.g.,][]{Brammer2012}.  
In such cases, multiple dispersion angles can break this degeneracy \citep[see, e.g.,][]{Naidu24_ALT}, or follow-up 
integral-field or slit spectroscopy may be required to secure redshift measurements and resolve ambiguities.

\cite{Xiao2024} presented a systematic study of the stellar masses of FRESCO emission-line galaxies within the first billion years after the Big Bang, reporting that the stellar masses of all galaxies are consistent with the $\Lambda$CDM model. However, three sources (i.e., S1, S2, and S3) were identified with unexpectedly high stellar masses and baryon-to-star conversion efficiencies.
Among these, the source S1 ($\alpha$, $\delta$ [J2000] = 53.120441, -27.742092) 
was reported at a spectroscopic redshift of $z_{\rm grism} = 5.58$, derived from JWST/NIRCam slitless grism observations from FRESCO \citep{Oesch2023}, and played a role in establishing the extreme nature of the sample.

S1 was originally identified in the GOODS-ALMA survey as a highly dust-obscured, optically dark galaxy, being undetected in \textit{HST} imaging while bright at submillimeter wavelengths \citep{Gomez-Guijarro2022a, Xiao2023}. It is robustly detected at 850\,$\mu$m with JCMT/SCUBA-2, at 870\,$\mu$m with ALMA (ID68 in \citealt{Cowie2018}), and at 1.13\,mm in GOODS-ALMA 2.0 \citep{Gomez-Guijarro2022a}. 
At shorter wavelengths, S1 is heavily blended with two nearby sources in the \textit{Spitzer}/IRAC 3.6 and 4.5\,$\mu$m bands, which led to its classification as an IRAC-dropout candidate \citep{Xiao2023}. 
Earlier, it has also been reported as a K-band dropout in the ASAGAO survey (ID20 in \citealt {yamaguchi2019}). 
Despite its bright dust emission, S1 remained undetected in the rest-frame optical until the FRESCO survey, which provided the first significant detection in this regime \citep{Xiao2024}.

In this work, we present new JWST/NIRSpec IFU spectroscopy of S1, which revises its redshift to $z_{\rm spec}=3.2439\pm 0.0002$.  
We demonstrate that the earlier grism-based redshift was driven by contamination from a nearby galaxy whose dispersed spectral trace overlaps with the one of S1.  
We also report new ALMA 1\,mm continuum measurements, update the UV-to-FIR SED of S1 at the revised redshift, and reassess its physical properties in light of these results.

This paper is organized as follows.  
In Section~\ref{Sec: data}, we describe the JWST/NIRSpec IFU observations, the ALMA band 7 observations, and the compiled multi-wavelength photometry.  
In Section~\ref{Sec: redshift}, we present the revised spectroscopic redshift from the IFU data and identify the origin of the previous grism-based misidentification.  
In Section~\ref{Sect:sed}, we update the UV-to-FIR SED modeling of S1 at the revised redshift and reassess its physical properties.  
In Section~\ref{Sect:discussion}, we discuss the nature of S1.
We summarize our main findings in Section~\ref{Sec:conclusion}.  

Throughout this paper, we adopt a Chabrier initial mass function \citep[IMF;][]{Chabrier2003} to estimate the SFR and stellar mass ($M_{\star}$). We assume a Planck cosmology \citep{Planck2020} with $(\Omega_\mathrm{m},\, \Omega_\mathrm{\Lambda},\, h,\, \sigma_\mathrm{8})=(0.3,\, 0.7,\, 0.7, \,0.81)$. When necessary, data from the literature have been converted with a conversion factor of $M_{\star}$ \citep[][IMF]{Salpeter1955} = 1.7  $\times$ $M_{\star}$ \citep[][IMF]{Chabrier2003}. All magnitudes are in the AB system \citep{Oke1983}, such that $m_{\rm AB} = 23.9 - 2.5$ $\times$ log(S$_{\nu}$ [$\mu$Jy]).

\section{Observations and Data}\label{Sec: data}

\subsection{JWST/NIRSpec IFU}\label{Sect:ifu} 
S1 was observed with JWST/NIRSpec in IFU mode using the G395H/F290LP configuration (Program ID: 5572; PI: M. Xiao).
This setup provides spectral coverage over 2.87-5.27\,$\mu$m at a resolving power of $R \sim 2700$, well suited for detecting rest-frame optical emission lines at $z \sim 3-6$. It also enables spatially resolved kinematic studies, which constitute a primary objective of this program.

The observations were obtained using the NRSIRS2 readout mode with a 16-point cycling dither pattern with medium offsets for 20 Groups.
For S1, the total on-source integration time was 8.9 hours.
No dedicated background exposures were taken, as the IFU field of view contains sufficient empty regions for the background subtraction.

The data were reduced using the \texttt{msaexp}\footnote{https://github.com/gbrammer/msaexp} software \citep{msaexp} that itself makes extensive use of the standard \texttt{jwst} pipeline, following the procedure described in \citet{Torralba2025}.
Detector-level processing was carried out using the \texttt{Detector1Pipeline}, including standard corrections for bias, dark current, flat-fielding, and cosmic rays.
Residual $1/f$ noise was mitigated on a per-detector basis, and additional corrections were applied to remove transient artifacts associated with cosmic-ray showers.

From the final calibrated exposures produced by the \texttt{Spec2Pipeline}, we extract pixel tables where each detector pixel within the 30 IFU slices is tagged with values of wavelength, spatial offset position relative to the telescope pointing, intensity, variance, and a data quality flag.  We add an additional data quality flag where any detector pixel that saturates in a particular exposure -- e.g., from a cosmic ray ``snowball'' \citep{Rigby2023} -- is flagged in subsequent exposures that will likely show some residual persistent signal unrelated to the science target.  Pixels masked by this saturation flag or the standard \textsc{do\_not\_use} data quality flag are excluded from subsequent analysis.

The pixel tables from the 16 dithered exposures are combined into a data cube where the pixels within a wavelength bin are resampled on a grid of square 0\farcs05 spatial pixels using an algorithm modeled after \texttt{drizzle} \citep{fruchter02}.  The output spatial pixel grid is set to be similar to the orthogonal axes along and across the IFU slices, such that the output grid is as similar as possible to the spatial sampling of the original detector pixels.  The spatial white-light image created from an inverse-variance weighted mean along the cube is shown in Fig.~\ref{fig_spec}.  The sky spectrum is extracted for spaxels outside of a source mask created from the white line image and subtracted from the data cube.  We extract a 1D spectrum (Fig.\ref{fig_spec}) of the primary target S1 with an optimal extraction \citep{horne1986} using the white light image within a radius of 5 pixels (0\farcs25) of the S1 centroid as the spatial template.

Emission lines were identified and fitted with Gaussian profiles to determine the spectroscopic redshift (Figs~\ref{fig_spec}, \ref{fig_zoom}; see details in Sect.~\ref{Sec: redshift1}).

\subsection{ALMA Observations}\label{Sect:alma} 
S1 was observed with ALMA as part of a Cycle~10 program designed to target the [C\,\textsc{ii}]~158\,$\mu$m emission line and the underlying 1mm dust continuum, assuming a spectroscopic redshift of $z = 5.579$ (Project ID: 2023.1.00837.S; PI: M.~Xiao).
The observations were carried out in Band~7 using a single pointing centered on the target coordinates. 
The total on-source integration time was 0.5\,hour.

The correlator setup was optimized to cover the expected [C\,\textsc{ii}] line at the assumed redshift, with one spectral window centered at 288.855\,GHz and three additional 2\,GHz-wide spectral windows placed in line-free regions for continuum measurements.
At the revised redshift of $z_{\rm spec} = 3.2439$ (see details in Sect.~\ref{Sec: redshift}), the [C\,\textsc{ii}] line falls outside the observed frequency range, and no other strong emission lines are expected to be covered and detected in the current dataset.

The continuum setup allows for a measurement of the dust emission.
Data calibration was performed using the standard ALMA pipeline within the Common Astronomy Software Application package \citep[CASA;][]{McMullin2007}, version 6.5.4.9.
Continuum imaging was carried out using the \textit{tclean} task, with a pixel scale of 0.1$^{\prime\prime}$ and a Briggs weighting with robust = 0.5. Compared to the use of a natural weighting that provides a better sensitivity, the current strategy can help improve the angular resolution.
The resulting continuum image reaches an rms sensitivity of $\sim$ 39\,$\mu$Jy\,beam$^{-1}$, with a synthesized beam of 0.618$^{\prime\prime} \times$ 0.491$^{\prime\prime}$ ($\sim$ 4.75 kpc $\times$ 3.78 kpc in physical scale).

We performed a two-dimensional elliptical Gaussian fit to the continuum map at the position of S1. The derived best-fit flux density is $S_{\rm 1mm} = 0.99\pm0.08$\,mJy, and half-light radius is $R_{\rm e, 1mm}=0.99\pm0.32$ kpc. To minimize potential uncertainties associated with the imaging process, we also fitted the visibility data directly in the \textit{uv} plane, with an elliptical Gaussian model (task \textit{uvmodelfit}). This yields a flux density of $S_{\rm 1mm} = 0.99\pm0.03$\,mJy and a half-light radius of $R_{\rm e, 1mm}=0.73\pm0.10$ kpc, consistent with the values derived from the image plane. We further use the values derived from the \textit{uv} plane throughout the whole paper.

In Fig.~\ref{fig_dust}, we show the dust distribution from ALMA at 1mm and the stellar emission from JWST. 
The inferred dust size of S1 at 1\,mm, if taken from the \textit{uv}-plane, is slightly smaller than its stellar size of $R_{\rm e, F444W} = 0.97\pm0.01$\,kpc measured from JWST/F444W imaging \citep{Xiao2024}. In addition, the dust size of S1 is consistent with the typical dust size of ALMA-detected OFGs at $z\sim4$ \cite[average $R_{\rm e}=0.80\pm0.03$ kpc;][]{Xiao2023} and dusty star-forming galaxies (DSFGs) at median $z\sim2.5$ from the GOODS-ALMA 1.13mm survey \cite[median $R_{\rm e}=0.73\pm0.29$ kpc;][]{Gomez-Guijarro2022a}. 

\subsection{Multi-wavelength photometry}\label{Sect:Photometry}
S1 is covered by extensive multi-wavelength observations spanning from X-ray to radio wavelengths, including \textit{Chandra}, \textit{HST}, \textit{JWST}, \textit{Spitzer}, \textit{Herschel}, SCUBA-2, ALMA, and the VLA.
Table~\ref{table:phot} summarizes the compiled photometry.
These measurements are used to perform updated UV-to-FIR SED fitting at the revised redshift.

\section{Revised Redshift and Origin of the Previous Misidentification}\label{Sec: redshift}

\subsection{Spectroscopic redshift from NIRSpec/IFU}\label{Sec: redshift1}
\begin{figure*}
\centering
\includegraphics[width=18cm]{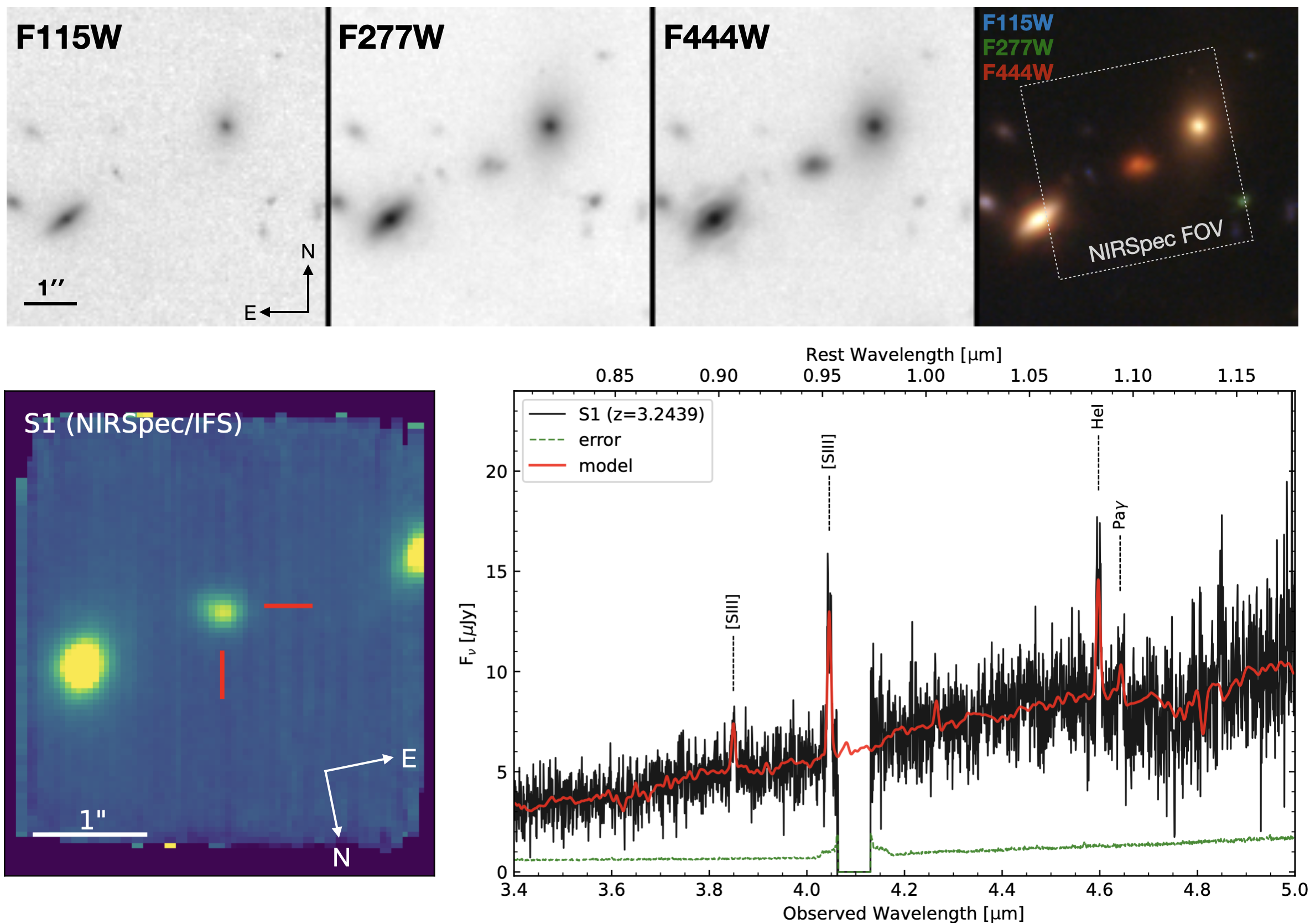}
\caption{Images and JWST/NIRSpec IFU spectrum of S1.
The top panels show image stamps in the F115W, F277W, and F444W filters, together with the corresponding RGB composite.
 The bottom panels show the IFU cube (left; continuum) and the extracted 1D spectrum (right). The data are shown in black, with 1$\sigma$ uncertainties in green, and the best-fit model is overplotted in red. 
Key emission lines used to determine the spectroscopic redshift are indicated.
 }
         \label{fig_spec}
\end{figure*}

The JWST/NIRSpec IFU spectrum of S1 reveals multiple rest-frame optical emission lines, including [S\,\textsc{iii}]$\lambda9069$, [S\,\textsc{iii}]$\lambda9531$, He\,\textsc{i}~$\lambda1.083\,\mu$m, and Pa$\gamma$ (Fig.~\ref{fig_spec}). 
We fit these features in the spatially integrated spectrum with a simple Gaussian model, adopting a representative intrinsic line width of $\sigma \simeq 250$\,km\,s$^{-1}$.
This yields a robust spectroscopic redshift of
\begin{equation}
z_{\rm spec} = 3.2439 \pm 0.0002.
\end{equation}

\begin{figure*}
\centering
\includegraphics[width=18cm]{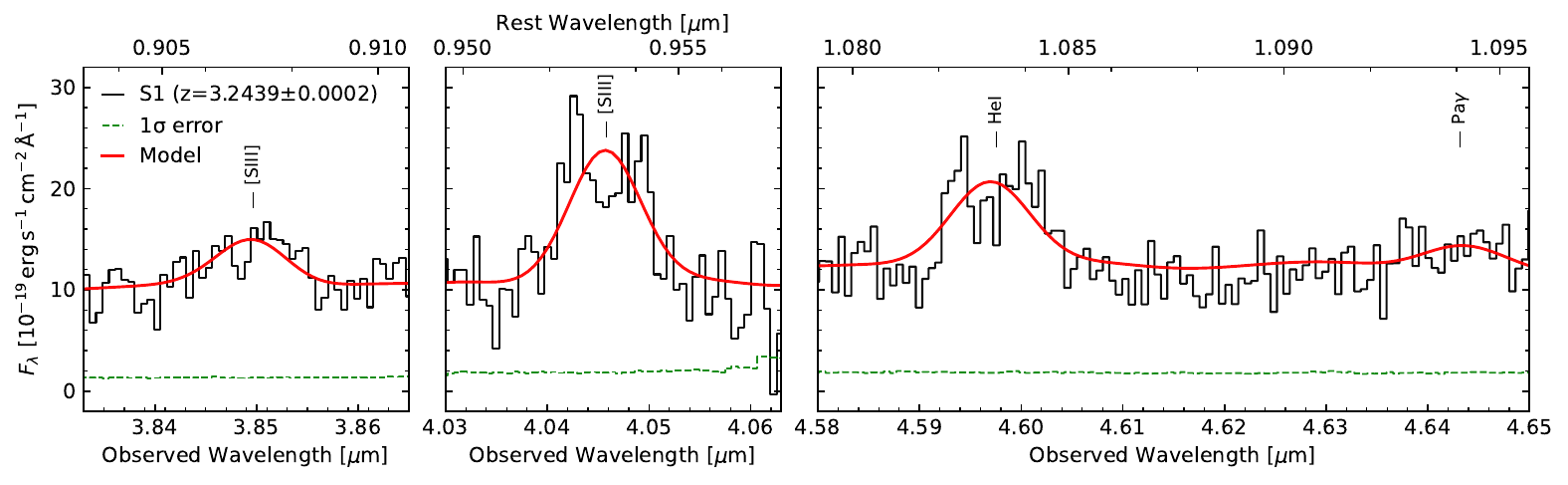}
\caption{Zoom-in spectra of [S\,\textsc{iii}]$\lambda9069$ (left), [S\,\textsc{iii}]$\lambda9531$ (middle), He\,\textsc{i}~$\lambda1.083\,\mu$m and Pa$\gamma$ (right).}
         \label{fig_zoom}
\end{figure*}

Both the [S\,\textsc{iii}]$\lambda9531$ and He\,\textsc{i} lines exhibit double-peaked profiles in the integrated spectrum.  
Zoom-in views of these features are shown in Fig.~\ref{fig_zoom}.  
The velocity separation between the blue and red peaks is $\sim$$100-200$~km~s$^{-1}$, suggesting that more than one kinematic component contributes to the emission. 

Such a double-peaked structure can arise from, e.g., internal velocity structure (e.g., rotation disk) and/or a minor merger. The precise choice of ``systemic'' redshift may vary slightly depending on whether one fits the full integrated profile or emphasizes one of the peaks.

\begin{figure*}
\centering
\includegraphics[width=15cm]{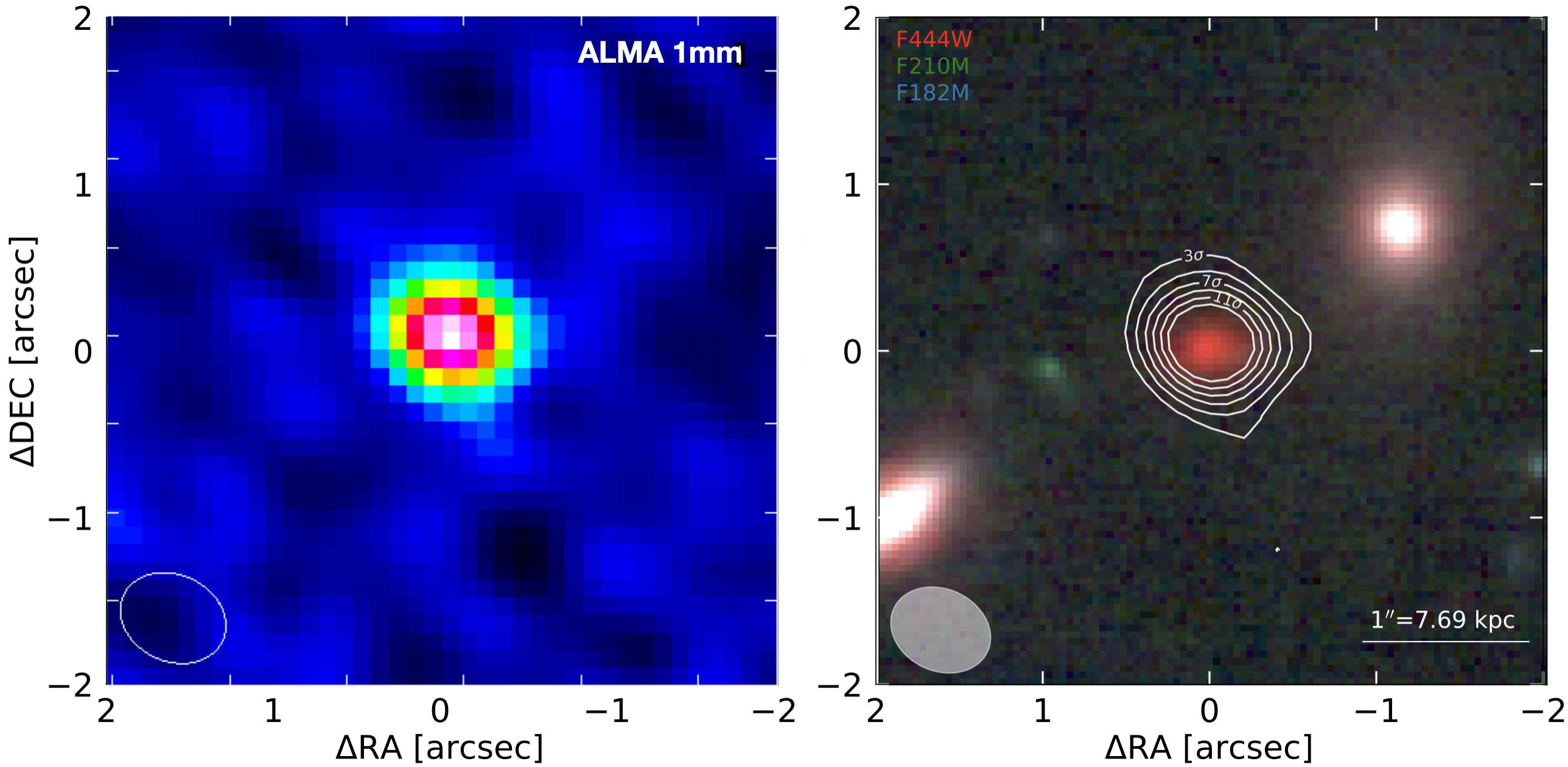}
\caption{Left: ALMA 1 mm dust continuum image of S1. The synthesized beam size (0.618$^{\prime\prime} \times$ 0.491$^{\prime\prime}$) is shown in the lower-left corner. Right: ALMA 1 mm emission contours overlaid on the JWST RGB image. Contour levels start at 3$\sigma$ and increase in steps of $\pm2\sigma$, with positive and negative contours shown as solid and dashed lines, respectively.  The intrinsic dust size ($R_{\rm e, 1mm}=0.73\pm0.10$ kpc) is slightly smaller than the stellar size from JWST F444W image \citep[R$_{\rm e, F444W}=0.97\pm0.01$ kpc;][]{Xiao2024}.}
         \label{fig_dust}
\end{figure*}

We note that, in the NIRCam imaging, S1 appears to show sub-structure at shorter wavelengths (e.g.,
F277W; Fig.~\ref{fig_spec}), which could indicate the presence of an additional component. 
However, S1 is subject to substantial dust obscuration ($A_V = 2.5\pm0.1$ mag; Table~\ref{table:sed}), and the ALMA 1\,mm continuum emission is centrally concentrated, tracing the main dust-enshrouded stellar body (Fig.~\ref{fig_dust}).  
The comparable half-light radii of the dust emission and the stellar light measured in F444W (Section~\ref{Sect:alma}) are consistent with a centrally obscured system.  
In this context, the apparent multi-component structure seen at shorter wavelengths could be naturally explained by differential or patchy dust attenuation, such that only part of the underlying stellar distribution is visible in F277W, and undetected in F115W.  

Taken together, the spectral and morphological evidence suggests that the multiple kinematic components are closely associated both spatially and in velocity space, and likely form a single physical system rather than distinct, widely separated galaxies.  
For this reason, we treat S1 as a single, coherent system and adopt the systemic redshift derived from fitting the full integrated profile.  

A full spatially resolved kinematic analysis to disentangle the merger versus rotating-disk scenarios is beyond the scope of this paper and will be presented in a companion study (Xiao et al., in prep.).  
In what follows, we fix $z_{\rm spec}=3.2439$ for all spectral and SED analyses of S1.


\begin{table}
\caption{Flux or flux densities of S1 from multi-wavelength observations.}   
\tiny          
\centering
\setlength{\tabcolsep}{0.6pt}
\renewcommand{\arraystretch}{1.2}
\begin{threeparttable}   
 
\begin{tabular}{l c c cc }    
\hline\hline       
                  
Instrument & $\lambda_{\rm ref}$ & Units & S1 & Ref.  \\ 
\hline  
\textit{Chandra}$^{a}$ & 0.5-7 keV & erg\,s$^{-1}$cm$^{-2}$ & $<5.7$$\times$$10^{-17}$ & \cite{Luo2017}\\
                 & 0.5-2 keV  & erg\,s$^{-1}$cm$^{-2}$ & $<1.9$$\times$$10^{-17}$ & \cite{Luo2017} \\
                 & 2-7 keV  & erg\,s$^{-1}$cm$^{-2}$ & $<8.1$$\times$$10^{-17}$ & \cite{Luo2017}\\
ACS F435W & 0.43\,\textmu m &  nJy & $<16.2$ &  \cite{Weibel2025_uvlf}\\
ACS F606W & 0.59\,\textmu m &  nJy & $<18.9$ &  \cite{Weibel2025_uvlf}\\
ACS F775W & 0.77\,\textmu m &  nJy & $<40.1$ &  \cite{Weibel2025_uvlf}\\
ACS F814W & 0.80\,\textmu m &  nJy & $<23.4$ & \cite{Weibel2025_uvlf}\\
ACS F850LP & 0.90\,\textmu m & nJy & $<36.9$ & \cite{Weibel2025_uvlf} \\
WFC3 F105W & 1.06\,\textmu m & nJy & $<221.5$ & \cite{Weibel2025_uvlf}\\
WFC3 F125W & 1.25\,\textmu m & nJy & $<27.0$& \cite{Weibel2025_uvlf} \\
WFC3 F140W & 1.40\,\textmu m & nJy &  $<108.6$& \cite{Weibel2025_uvlf}\\
WFC3 F160W & 1.54\,\textmu m & nJy & 40.1 $\pm$ 11.3 & \cite{Weibel2025_uvlf} \\

NIRCam F090W$^{b}$ & 0.90\,\textmu m  & nJy &  $<41.0$ & \cite{Lapasia2026} \\
NIRCam F115W & 1.15\,\textmu m  & nJy & $<35.4$ & \cite{Lapasia2026}  \\
NIRCam F150W &  1.50\,\textmu m & nJy & 89.3 $\pm$ 15.4 & \cite{Lapasia2026}  \\
NIRCam F182M &  1.85\,\textmu m & nJy & 278.7 $\pm$ 19.4 & \cite{Lapasia2026} \\
NIRCam F200W & 1.99\,\textmu m  & nJy & 227.0 $\pm$ 11.3 & \cite{Lapasia2026} \\
NIRCam F210M &  2.10\,\textmu m & nJy & 282.6 $\pm$ 26.8 & \cite{Lapasia2026}  \\
NIRCam F277W &  2.77\,\textmu m & nJy & 701.6  $\pm$ 35.1 & \cite{Lapasia2026} \\
NIRCam F335M &  3.35\,\textmu m & nJy & 916.9 $\pm$ 45.8 & \cite{Lapasia2026}  \\
NIRCam F356W &  3.57\,\textmu m & nJy & 1131.2 $\pm$ 56.6 & \cite{Lapasia2026}  \\
NIRCam F410M & 4.10\,\textmu m  & nJy & 1713.3 $\pm$ 85.7 & \cite{Lapasia2026}  \\
NIRCam F444W & 4.40\,\textmu m  & nJy & 2052.1 $\pm$ 102.6 & \cite{Lapasia2026}  \\

\textit{Spitzer}/MIPS$^{c}$ & 24\,\textmu m &  \textmu Jy & $<20 $ & \cite{Elbaz2011}\\
                      & 70\,\textmu m &  mJy & $<3.2 $ & \cite{Elbaz2011}  \\
\textit{Herschel}/PACS$^{d}$ & 100\,\textmu m &  mJy & $<49.7$ & \cite{Leroy2024} \\
                       & 160\,\textmu m &  mJy & $<63.8$ & \cite{Leroy2024} \\
\textit{Herschel}/SPIRE$^{d}$ & 250\,\textmu m & mJy & $<23.0$ & \cite{Leroy2024} \\
                        & 350\,\textmu m & mJy & $<22.4$ & \cite{Leroy2024}\\
                        & 500\,\textmu m & mJy & $<19.7$ & \cite{Leroy2024} \\
 SCUBA-2 & 850\,\textmu m & mJy & $2.76 \pm 0.47$ & \cite{Cowie2018} \\
ALMA & 870\,\textmu m & mJy & $1.35 \pm 0.24$ & \cite{Cowie2018}  \\
 & 1\,mm &  mJy & $0.99 \pm 0.03$ & This work \\
 & 1.13\,mm & mJy & $0.95 \pm 0.12$  & \cite{gomez2021}\\
VLA$^{e}$ & 5\,cm  & uJy & $<0.9$ & \cite{Rujopakarn2016}  \\
& 10\,cm  & uJy & $<2.5$ & \cite{Rujopakarn2016} \\

\hline 
                 
\end{tabular}
\begin{tablenotes}
\item \textbf{Note:} We use 1$\sigma$ for uncertainties and 3$\sigma$ upper limits for non-detections. (a) We calculate 3$\sigma$ upper limits based on the mean sensitivity limits achieved in the central $\sim$1 arcmin$^2$ area at the average aim point from the 7 Ms Chandra Deep Field-North survey \citep{Luo2017}. 
(b) We adopt the same values for all the NIRCam photometry of S1 from \cite{Lapasia2026} to keep consistency.
(c) and (d)  Sources are not detected, we adopt 3$\sigma$ upper limits from the GOODS-Herschel survey \citep{Elbaz2011} and from \cite{Leroy2024}, respectively. 
(e) VLA image at 6 GHz (5 cm) and 3 GHz (10 cm) \citep[PI: W. Rujopakarn, private communication;][]{Rujopakarn2016}.
\end{tablenotes}
\label{table:phot} 
\end{threeparttable} 
\end{table}


\begin{figure*}
\centering
\includegraphics[width=18.5cm]{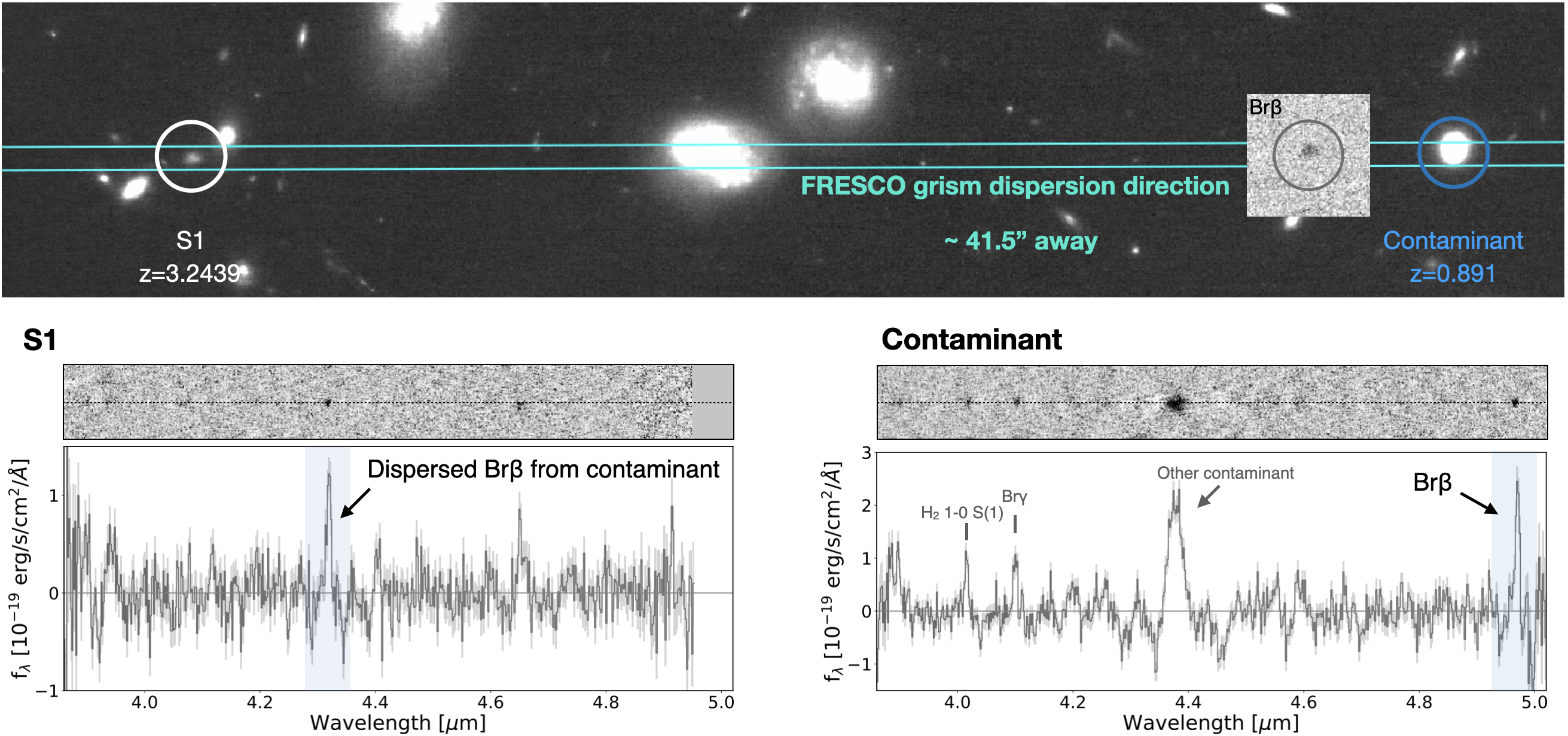}
\caption{Illustration of the slitless grism contamination affecting the original redshift estimate of S1.
The cyan lines in the top panel indicate the dispersion direction of the FRESCO grism.  
A nearby galaxy at $z=0.891$ \citep[ID CDFS\_00336 in][]{Mignoli2005}, located $\sim41.5^{\prime\prime}$ away from S1 along the dispersion axis, produces a Br$\beta$ emission feature (bottom right; see also the Br$\beta$ map inset in the top panel) whose spectral trace overlaps with that of S1 in the slitless data. This overlap led to the earlier misidentification of the H$\alpha$ line as being associated with S1 (bottom left). The bottom panels show the 2D and 1D FRESCO NIRCam/grism spectra (F444W filter) of S1 and the contaminant, respectively.
}
         \label{fig_contaminant}
\end{figure*}

For completeness, we note that the same JWST/NIRSpec IFU program (Program ID: 5572; PI: M. Xiao) also includes observations of another source from the ultra-massive galaxy sample presented in \citet{Xiao2024}, namely S3.
A preliminary inspection of the IFU data confirms a spectroscopic redshift of $z=5.1802$ for S3, fully consistent with the redshift reported in \citet{Xiao2024}.
A detailed kinematic analysis of S3, together with that of S1, will be presented in Xiao et al. (in prep.).

\subsection{Origin of the Previous Misidentification}\label{Sect:comtaminant}

In \citet{Xiao2024}, we reported a single high-significance (S/N $>14$) emission feature at $\lambda_{\rm obs}\simeq4.32\,\mu$m in the FRESCO NIRCam slitless grism spectrum of S1.  
Given the lack of additional lines, we carried out an extensive analysis of possible line identifications and redshift solutions, considering H$\alpha$, Pa$\alpha$, Pa$\beta$, and other nebular transitions.  After considering the agreement with the photometric redshift, the constraints from infrared properties, and the similar morphology of the emission line and S1 in F444W (Fig.~\ref{fig_contaminant}), we ultimately favored the interpretation of the feature as H$\alpha$ at $z_{\rm grism}=5.58$.

However, the deep JWST/NIRSpec IFU spectrum shows no emission line at the wavelength expected at $\lambda_{\rm obs}\simeq4.32\,\mu$m (Fig.~\ref{fig_spec}), even though such a line would have been easily detected at the sensitivity of the IFU data.  
This demonstrates that the previously detected grism feature does not originate from S1 itself.  
Therefore, the key issue is not the line identification, but rather the source attribution in the slitless data.

To investigate the origin of the grism feature, we re-examined the two-dimensional (2D) grism spectra and associated trace diagnostics.  
As shown in Fig.~\ref{fig_contaminant}, the emission feature previously attributed to S1 is most plausibly explained by contamination from a nearby galaxy ($\alpha,\delta$ [J2000] = 53.1275129, -27.7420060) at $z=0.891$ \citep[ID CDFS\_00336;][]{Mignoli2005}, whose dispersed spectral trace overlaps with that of S1 along the dispersion direction.  
In particular, the Br$\beta$ emission line from this source intersects precisely with the location of the spurious feature in the S1 grism spectrum.
This accounts for both the wavelength and spatial location of the previously reported line.

This case illustrates a known challenge of slitless spectroscopy: when only a single emission feature is detected, overlapping spectral traces can complicate source attribution.  
While grism surveys are extremely powerful for blind line searches \citep[e.g.,][]{Oesch2023, Meyer2025_cosmos3d}, such ambiguities underscore the importance of IFU, slit spectroscopy, or multiple line identification for secure redshift confirmation.  
At the same time, it is important to note that several observational strategies can substantially mitigate trace-overlap ambiguities, including observations at multiple position angles (e.g., ALT; \citealt{Naidu24_ALT} and the forthcoming TWINKLE program; \citealt{Naidu2025_twinkle}) and the use of multiple grism filters (e.g., MAGNIF; \citealt{Sun2023_MAGNIF}), which improve the identification and rejection of contaminants in dense or lensed fields.

Our analysis also highlights practical diagnostics that can reduce the risk of misattribution in single-line cases:  
($i$) checking whether the same line solution is simultaneously favored for multiple sources aligned along the dispersion axis;  
($ii$) explicitly inspecting all potential contaminants whose traces intersect the target spectrum at the relevant wavelength; and  
($iii$) assessing the physical plausibility of the inferred line identification (e.g., whether an apparently isolated H$\alpha$ detection without accompanying \nii\ or \sii\ emission is expected for the source and at the achieved sensitivity).  
Applying such checks can motivate more conservative redshift-quality flags or help prioritize follow-up spectroscopy when ambiguities remain.

We therefore conclude that the earlier $z_{\rm grism}=5.58$ solution for S1 was caused by contamination in the slitless grism data rather than a genuine emission line from S1 itself.

\section{Updated Physical Properties from SED}\label{Sect:sed}

We perform new SED fitting using the revised redshift and the more complete multi-wavelength photometry with CIGALE \citep{Boquien2019}. In analogy to \citet{Xiao2024}, during the SED modeling, we apply three different parametric star formation histories (SFH), i.e., delayed-$\tau$ SFH (DSFH), DSFH + recent burst with constant SFR (DSFH+burst), and constant SFH (CSFH), to check the consistency and robustness of the physical properties derived. The nebular emission that could contribute to the broad and medium band photometry is also taken into account in the CIGALE modeling. As for the dust component, we apply an attenuation curve from \citet{Calzetti2000} and fit the emission with the dl2014 model \citep{Draine2014}. As the source remains undetected in both X-ray and mid-IR (see Table.~\ref{table:phot}), we do not include an AGN model in the SED fitting. The exact model of SED compositions applied, as well as their parameter ranges chosen in the CIGALE modeling, are summarized in Table\,~\ref{table:sedparam}.

\begin{table}
\centering
\caption{CIGALE models and parameters used to fit the SED of S1.}
\begin{tabular}{lll}
\hline
Model & Parameter & Values/Ranges \\
\hline
SFH (DSFH) & \texttt{tau\_main} (Myr) & [100, 1500] \\
sfhdelayed & \texttt{age\_main} (Myr) & [50, 1500] \\
           & \texttt{f\_burst} & 0 \\
SFH (CSFH) & \texttt{tau\_main} (Myr) & $10^6$ \\
sfh2exp    & \texttt{age\_main} (Myr) & [10, 1500] \\
           & \texttt{f\_burst} & 0 \\
SFH (DSFH+burst) & \texttt{tau\_main} (Myr) & [100, 500] \\
sfhdelayed & \texttt{age\_main} (Myr) & [500, 1500] \\
           & \texttt{tau\_burst} (Myr) & $10^6$ \\
           & \texttt{age\_burst} (Myr) & 10, 30, 100 \\
           & \texttt{f\_burst} & 0.0, [0.001, 0.3] \\           
\hline
nebular emission & \texttt{logU} & [-4.0,-1.0] \\
nebular & \texttt{zgas} & 0.02 \\
\hline
attenuation curve & \texttt{E\_BV\_lines} (mag) & [0.0, 2.0] \\
modified\_starburst & \texttt{E\_BV\_factor} & 1.0 \\
                           & \texttt{powerlaw\_slope} & 0.0 \\
\hline
dust emission & \texttt{qpah} & 2.5 \\
dl2014        & \texttt{umin} & [1.0, 50.0] \\
              & \texttt{gamma} & [0.001,0.99] \\
\hline              
\label{table:sedparam} 
\end{tabular}
\end{table}

\begin{table}
\centering
\caption{Updated Physical Properties of S1 at $z_{\rm spec}=3.2439$.}
\label{tab:sed}
\begin{tabular}{ccccc}
\hline
Parameter & Unit & CSFH & DSFH & DSFH+burst \\
\hline
$M_\ast$ & $10^{10}M_\odot$ & 3.6$\pm$0.7 & 3.7$\pm$0.9 & 4.0$\pm$0.8  \\
SFR & $M_\odot\,\mathrm{yr}^{-1}$ & 72$\pm$10 & 70$\pm$18 &  58$\pm$20  \\
$A_V$ & mag & 2.5$\pm$0.1 & 2.5$\pm$0.1 & 2.5$\pm$0.1  \\
$L_{\mathrm{IR}}$ & $10^{11}L_\odot$ & 8.2$\pm$0.8 & 7.8$\pm$1.4 & 7.3$\pm$1.6 \\
$M_{\rm dust}$ & $10^{9}M_\odot$ & 1.1$\pm$0.4 & 1.2$\pm$0.5 & 1.3$\pm$0.5 \\
\hline
\label{table:sed} 
\end{tabular}
\end{table}

\begin{figure}
\centering
\includegraphics[width=\linewidth]{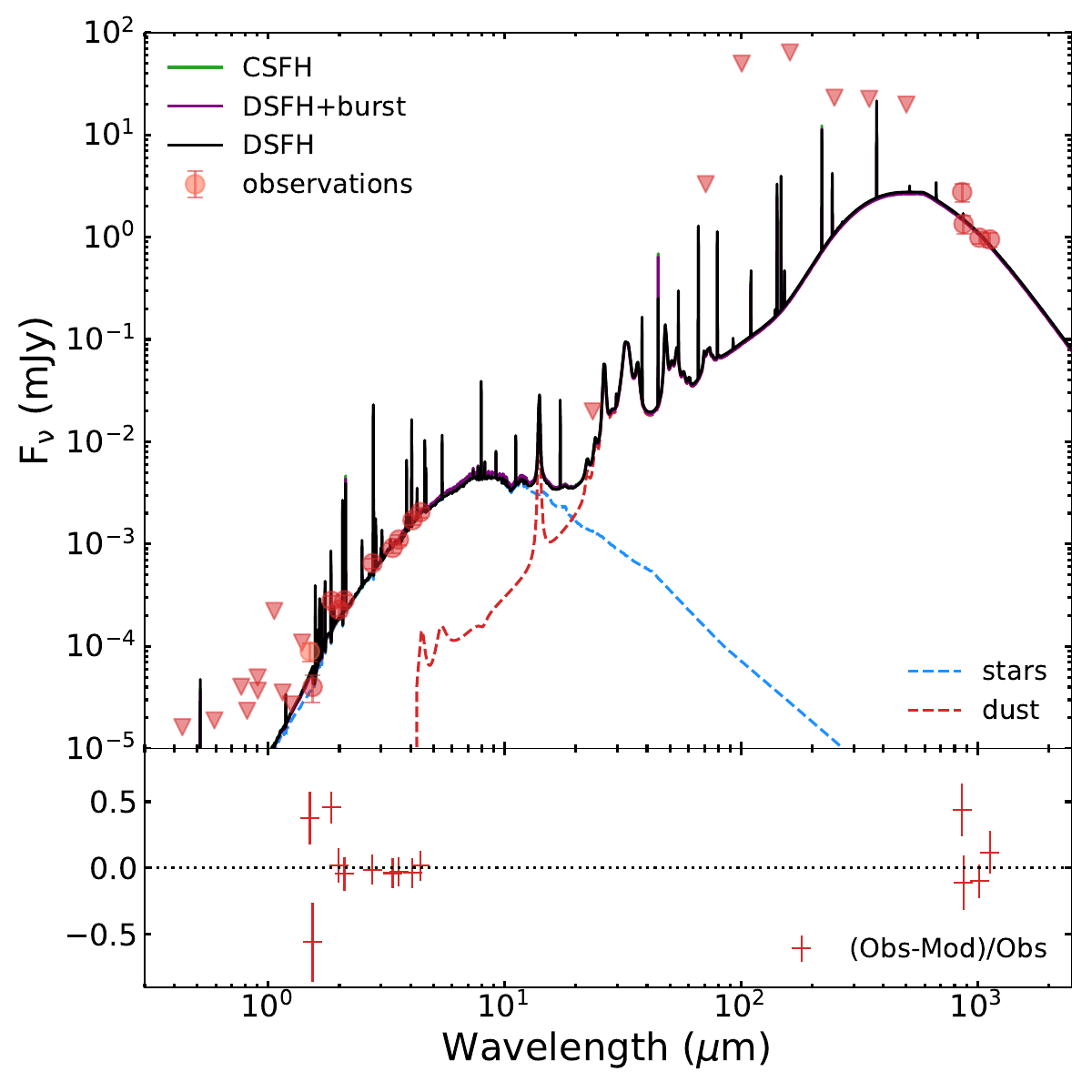}
\caption{CIGALE SED fitting on the multi-wavelength photometry of S1. The best-fit SED model under CSFH, DSFH, and DSFH+burst are plotted simultaneously in the top panel, but are almost indistinguishable from each other. Red and blue dashed lines display the best-fit stellar and dust components under DSFH, respectively. Fluxes and 3-$\sigma$ upper limits are shown with the red circles and triangles, respectively. The bottom panel shows the residual between the photometry and the best-fit model under DSFH. Note that the flux density at 850 $\mu$m might be boosted by the neighboring sources due to the large SCUBA-2 beam size.
}
\label{fig_sed}
\end{figure}

The updated physical parameters under the three different SFH models are summarized in Table~\ref{table:sed}. Compared to the original estimates at $z_{\rm grism} = 5.58$, our new fitting at $z_{\rm spec}=3.2439$ yields a $\sim6$ times lower $M_\star$, as well as a $\sim9$ times lower infrared luminosity ($L_{\rm IR}$) and SFR for S1, reflecting the reduced luminosity distance. The resulting dust attenuation in the rest-frame V band ($A_V$) is also slightly reduced compared to the previous estimation, but it remains significant and is consistent with typical dusty star-forming galaxies in the early Universe. The $M_\star$, SFR, $L_{\rm IR}$, and $A_V$ measurements are not sensitive to different SFH-parametrizations applied in our CIGALE modeling, as indicated in Table~\ref{tab:sed}. This is also illustrated in Fig.~\ref{fig_sed}, where the best-fit models to the multiwavelength photometry under three different SFHs are almost indistinguishable from each other. The derived star formation rate is consistent with \citet{Lapasia2026}, which applies a variable slope on the \citet{Calzetti2000} dust attenuation curve and non-parametric SFH with constant and rising SFH priors of S1 at the same redshift. For the stellar mass, we find that our measurements fall between the estimations from \citet{Lapasia2026} with constant and rising SFH priors, while the results are still consistent with both within the 1$\sigma$ uncertainties. Although our results presented in Table~\ref{table:sed} are derived using an attenuation curve fixed to \citet{Calzetti2000}, we notice that \citet{Lapasia2026} obtains a best-fit attenuation curve slope on S1 close to the standard \citet{Calzetti2000} value. Our test under the same CIGALE configuration, with a variable attenuation curve slope, also yields best-fit $M_\star$ and other parameters that are consistent with the results in Table~\ref{table:sed} within $1\sigma$.   

\section{Implications for the Nature of S1}\label{Sect:discussion} 

\begin{figure}
\centering
\includegraphics[width=\linewidth]{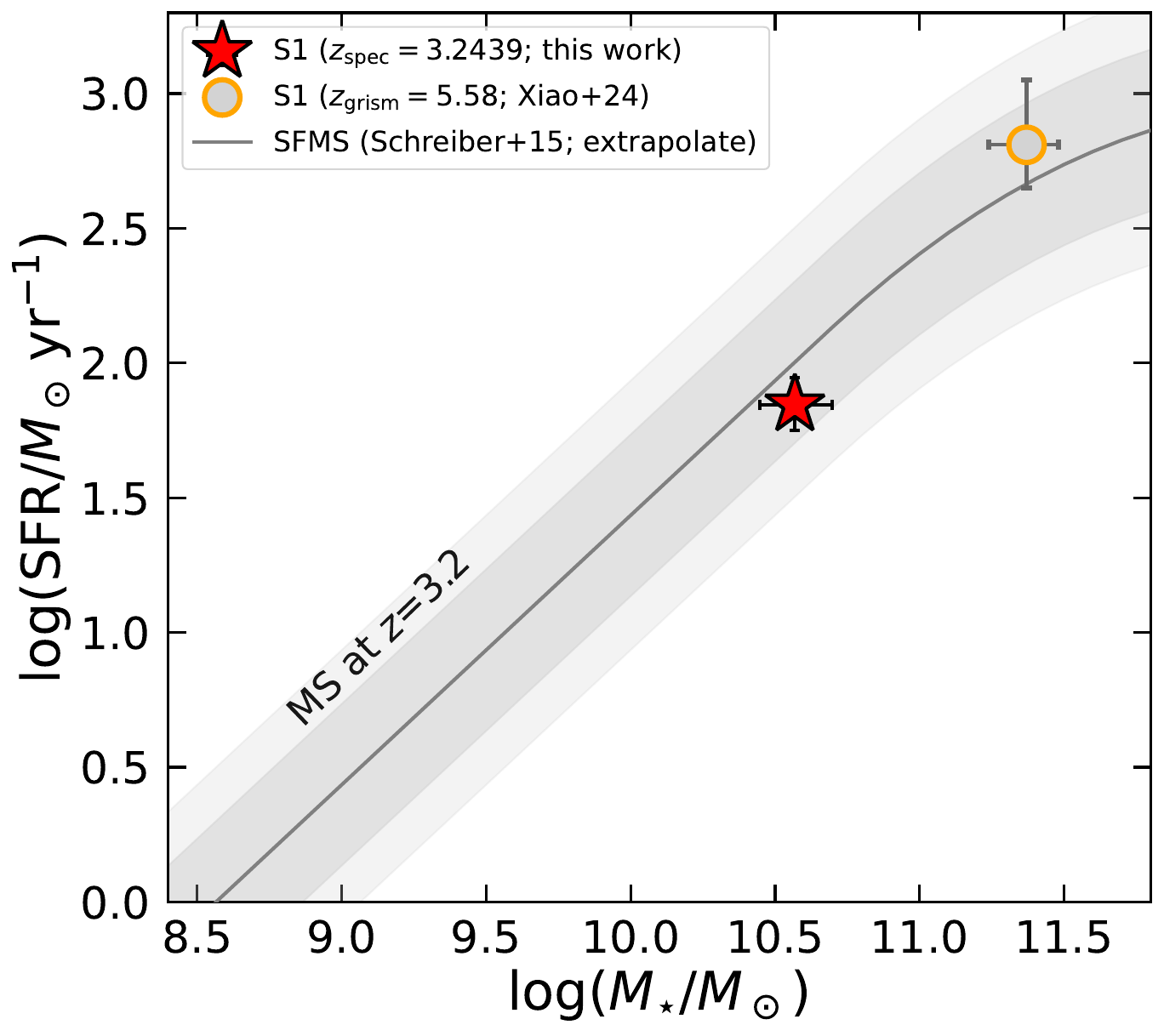}
\caption{Location of S1 (red star) compared to the SFMS in the SFR-$M_{\star}$ plane.  The grey-filled orange circles show the previous location of S1 in \citep{Xiao2024}.
Error bars correspond to 1$\sigma$ uncertainties.  The \cite{Schreiber2015} SFMS at $z=3.2$, 1$\sigma$ scatter ($\sim$0.3 dex), and a more extended typical scatter of $0.5$ dex are highlighted with a grey line, a dark grey shaded area, and a light grey shaded area, respectively. 
}
\label{fig_ms}
\end{figure}

The revised SED modeling with the updated redshift shows that S1 no longer belongs to the population of ultra-massive galaxies at $z \gtrsim 5$, so that its stellar mass and redshift can be reconciled with typical baryon-to-star conversion efficiencies in galaxy formation and evolution models. 
Nevertheless, S1 remains a massive galaxy. The derived $M_\star$ of S1 is comparable to the characteristic mass in the stellar mass function from COSMOS2020 \citep{Weaver2023} and recent JWST observations \citep[e.g.,][]{Weibel2024,Shuntov2025}. The derived SFR suggests S1 as a moderate star-forming system in the early Universe. As shown in Fig.~\ref{fig_ms}, S1 lies on the star-formation main sequence (SFMS) at its redshift \citep[e.g.,][]{Schreiber2015}.  
With a robust spectroscopic redshift and extensive multi-wavelength coverage, S1 therefore remains a valuable case for studying the evolution of massive galaxies across cosmic time.

The revised SED fitting further implies that S1 hosts a substantial dust reservoir, with a dust mass of $M_{\rm dust}\sim10^9\,M_\odot$. Combined with the inferred stellar mass, this corresponds to a dust-to-stellar mass ratio of $M_{\rm dust}/M_\star\sim3.2\%$. While this ratio exceeds the typical average values ($M_{\rm dust}/M_\star\sim 0.5-1.0\%$) reported for bright $z=2-5$ SMGs \citep{dacunha2015}, ALMA-selected $z\sim2.5$ DSFGs \citep{Gomez-Guijarro2022a}, and $z=2.0-2.5$ normal SFGs \citep[]{Shivaei2022}, such richness of dust content is not unexpected given the large intrinsic scatter observed among gas-rich galaxies at high redshift \citep[e.g.,][]{Dessauges-Zavadsky2025}.

A large dust mass naturally implies a significant cold gas reservoir, although the inferred gas mass depends on the assumed gas-to-dust ratio (GDR). In our estimation, we adopt a GDR of 90, consistent with the average of massive, chemically evolved ALESS SMGs \citep{swinbank2014}. Higher median GDR values of $\sim 300-500$ are commonly reported for $z=2-5$ star-forming galaxies \citep[e.g.,][]{Sawant2025}, but these systems typically have stellar masses an order of magnitude lower than that of S1. Previous studies have established a negative correlation between metallicity and GDR \citep[e.g.,][]{remy-ruyer2014,Popping2023}. Given its high stellar mass, large dust attenuation, and dust-rich nature, S1 is therefore more likely to be chemically mature, with a GDR closer to that of the ALESS SMG population than to the lower-mass high-$z$ galaxies studied by \citet{Sawant2025}.

Even under this relatively conservative assumption for the GDR, the inferred cold gas mass of S1 remains very high, with $M_{\rm gas} \sim 10^{11}\,M_\odot$. When combined with its moderate SFR of $\sim 70\,M_\odot/\rm yr$, this implies a long gas depletion timescale of $\tau_{\rm dep}=M_{\rm gas}/\rm SFR \sim 1.4$ Gyr, indicating a low star-formation efficiency. This value is significantly longer than the short depletion timescales of the typical $z\sim2-5$ SMGs \citep[$\tau_{\rm dep}\sim 100-200$ Myr;][]{Liao2024} and $z\sim4$ OFGs \citep[median $\tau_{\rm dep}$ =  $110^{+29}_{-30}$ Myr in][]{Xiao2023}, and even exceeds the typical values found for normal star-forming galaxies at similar redshifts \citep[$\tau_{\rm dep} \sim 300-700$ Myr;][]{tacconi2018,WangT2022}. 

Taken together, the combination of a large gas reservoir and a relatively low star-formation efficiency (high $\tau_{\rm dep}$) indicates that S1 is distinct from compact, merger-driven starbursts that rapidly exhaust their fuel and quench on short timescales.  
Instead, S1 appears to sustain its growth through a more gradual mode of star formation, converting gas into stars even more gently than the average main-sequence galaxy.  
These properties are consistent with a scenario in which its evolution is regulated by a quasi-equilibrium between gas inflow and star formation \citep{lilly2013, tacconi2020, aravena2020}.

\section{Conclusions}
\label{Sec:conclusion}

We present new JWST/NIRSpec IFU spectroscopy of the dusty massive galaxy S1, revising its redshift from the NIRCam grism-based value $z_{\rm grism}=5.58$ to a secure spectroscopic redshift of $z_{\rm spec}=3.2439\pm0.0002$. The IFU spectrum reveals multiple emission lines ([S\,\textsc{iii}]$\lambda9069$, [S\,\textsc{iii}]$\lambda9531$, He\,\textsc{i}~$\lambda1.083\,\mu$m, and Pa$\gamma$), from which we derive the systemic redshift by fitting the spatially integrated line profiles. The strongest lines show mildly double-peaked profiles in the integrated spectrum, suggesting potentially complex internal kinematics and/or multiple components; a full spatially resolved kinematic interpretation requires dedicated analysis and will be presented in future work (Xiao et al., in prep.).

We identify the origin of the previous grism-based misidentification. The single emission feature at $\lambda_{\rm obs}\simeq4.32\,\mu$m reported in \citet{Xiao2024} does not originate from S1, but instead arises from contamination by a nearby galaxy \citep[at $z=0.891$; ID CDFS\_00336 in][]{Mignoli2005}, whose dispersed spectral trace overlaps with that of S1 in the FRESCO slitless grism data. While slitless spectroscopy is a powerful and highly efficient tool for blind emission-line searches over wide fields, this case highlights an important caveat: when only a single feature is detected, overlapping spectral traces can lead to ambiguous source attribution and, consequently, to incorrect redshift assignments.

The revised redshift also clarifies the interpretation of the ALMA Cycle~10 follow-up observations originally designed to target [C\,\textsc{ii}]~158\,$\mu$m at $z_{\rm grism}=5.579$. At $z_{\rm spec}=3.2439$, the [C\,\textsc{ii}] line falls outside the observed frequency coverage, naturally explaining the lack of a line detection. Nevertheless, we robustly detect the 1\,mm dust continuum with a flux density $S_{\rm 1mm}=0.99\pm0.03$\,mJy, and infer a dust size with a half-light radius of $R_{\rm e, 1mm}=0.73\pm0.10$\,kpc, slightly smaller than the stellar size measured in JWST/F444W (R$_{\rm e, F444W}=0.97\pm0.01$\,kpc).

Finally, we update the UV-to-FIR SED modeling of S1 using the revised redshift and the compiled multi-wavelength photometry. The updated physical properties indicate that S1 is less extreme than previously inferred at $z_{\rm grism}=5.579$, but remains a massive (log$M_{\star}/M_{\odot}$ $\sim10.6$), heavily obscured ($A_V = 2.5\pm0.1$ mag), dust- and gas-rich ($M_{\rm dust}\sim10^9\,M_\odot$ and $M_{\rm gas} \sim 10^{11}\,M_\odot$) star-forming galaxy at $z_{\rm spec}=3.2439$. In particular, the inferred long depletion time ($\tau_{\rm dep}\sim1.4$\,Gyr) distinguishes S1 from high-$z$ SMGs and OFGs and suggests comparatively inefficient star formation, even relative to typical massive MS galaxies at similar redshift. With a secure spectroscopic redshift and extensive panchromatic constraints, S1 therefore remains a valuable case for future studies of dusty massive-galaxy evolution.

With respect to the broader interpretation of \citet{Xiao2024}, we conclude that the revised redshift of S1 reduces the number of ultra-massive galaxies, but does not alter the main conclusions of that work. The two key results remain: first, none of the galaxies in that study fundamentally challenge the $\Lambda$CDM framework. Second, even after removing S1, ultra-massive systems (e.g., S2) still require $\sim50\%$ of the baryons in their halos to be converted into stars, demonstrating the rapid assembly of massive galaxies in the early Universe.

More broadly, this work emphasizes that redshifts based on single-line detections from slitless spectroscopy should be treated with caution, particularly for sources that are rare and scientifically influential. IFU or slit spectroscopy mitigates this ambiguity by providing spatially resolved information and multi-line confirmation, enabling secure redshift measurements and reliable interpretation of follow-up observations.

\begin{acknowledgements}
This work is based in part on observations made with the NASA/ESA/CSA James Webb Space Telescope. The data were obtained from the Mikulski Archive for Space Telescopes at the Space Telescope Science Institute, which is operated by the Association of Universities for Research in Astronomy, Inc., under NASA contract NAS 5-03127 for JWST. These observations are associated with program 5572. Support for program JWST-GO-5572 was provided by NASA through a grant from the Space Telescope Science Institute, which is operated by the Association of Universities for Research in Astronomy, Inc., under NASA contract NAS 5-03127. 
This paper makes use of the following ALMA data: ADS/JAO.ALMA\#2023.1.00837.S. ALMA is a partnership of ESO (representing its member states), NSF (USA) and NINS (Japan), together with NRC (Canada), NSTC and ASIAA (Taiwan), and KASI (Republic of Korea), in cooperation with the Republic of Chile. The Joint ALMA Observatory is operated by ESO, AUI/NRAO and NAOJ. The National Radio Astronomy Observatory is a facility of the National Science Foundation operated under cooperative agreement by Associated Universities, Inc.
This work has received funding from the Swiss State Secretariat for Education, Research and Innovation (SERI) under contract number MB22.00072, as well as from the Swiss National Science Foundation (SNSF) through project grant 200020\_207349. 
D.M. acknowledges support for program JWST-GO-01895, provided by NASA through a grant from the Space Telescope Space Institute, which is operated by the Associations of Universities for Research in Astronomy, Incorporated, under NASA contract NAS5-26555.
\end{acknowledgements}

\bibliography{reference_update}{}
\bibliographystyle{aa}

\end{document}